# Open-source Software and Pipelines for Generalizable and Reproducible EEG Analysis: An Overview

**Takfarinas Medani[1], Raymundo Cassani[2]**

[1]University of Southern California [2]McGill University

## 1 Reproducibility, Replicability, Robustness, and Generalizability

In scientific research, the concepts of **reproducibility**, **replicability**, **robustness**, and **generalizability** are often misunderstood or used inconsistently. In the following section, we aim to define them and describe their relationship in the context of EEG data analysis.

*Reproducibility* is the ability to obtain the same results using the original data and analysis procedures. Essentially, if one has access to the EEG data and the methods (usually in the form of code pipelines) used to analyze it, as described in a published work, one should be able to replicate the published findings [1, 2, 3, 4, 5, 6, 7]. In contrast, *replicability* refers to achieving consistent results when applying the same methods to new EEG data (e.g., a new sample or a repeated experiment), thereby demonstrating that the findings are not specific to a single dataset. *Robustness* refers to the stability of analysis results against reasonable variations in analysis method choices; for example, whether a result remains consistent across different preprocessing parameters or analytic pipelines. In practice, these three can diverge: many researchers may "reproduce" an EEG study's analysis using the shared data/code but struggle to *replicate* the effect on their own data, or find that results are not *robust* to small changes in filtering or artifact removal. A broader term is **generalizability**, which further extends the previous concepts by assessing whether findings remain consistent across both independent datasets and reasonable variations in analysis pipelines, experimental paradigms, populations, or recording conditions, thereby indicating that observed EEG effects reflect underlying neurophysiological principles rather than dataset- or method-specific factors. Ensuring that all concepts—including reproducibility, replicability, robustness, and generalizability—have become a cornerstone concern in EEG research [1, 2, 3, 5, 6, 8, 9]. Improving these aspects requires standardized practices and transparent reporting in EEG and MEG studies.

Table 1: Conceptual distinctions between reproducibility, replicability, robustness, and generalizability in data analysis pipelines. Adapted from the framework introduced by Nichols et al. (2022) and reproduced from the original publication in *NeuroImage* [3].

| Category | Data | Analysis | Primary Goal ("Why") |
|---|---|---|---|
| **Reproducible** | Same ($data_1$) | Same ($analysis_1$) | **Verify Integrity:** To ensure the original code/math produces the stated results. |
| **Replicable** | Multiple/New ($data_1$, $data_2$) | Same ($analysis_1$) | **Verify Reliability:** To ensure the finding isn't a fluke of one specific sample. |
| **Robust** | Same ($data_1$) | Multiple/New ($analysis_1$, $analysis_2$) | **Verify Methodology:** To ensure the result doesn't change if you use a different statistical tool. |
| **Generalizable** | Multiple/New ($data_1$, $data_2$) | Multiple/New ($analysis_1$, $analysis_2$) | **Verify Universality:** To ensure the finding holds across different groups and methods. |

# 2 Why Generalizability Is Challenging

EEG research faces unique hurdles to being generalizable. These challenges have their origins in two broad, interacting sources: heterogeneity in EEG data and in analysis methods and tools.

## 2.1 Heterogeneity of EEG Datasets

EEG datasets vary in several critical respects. First, there is *variability in hardware and file formats*: EEG systems are produced by dozens of manufacturers with different amplifier technologies, electrode types, reference schemes, and proprietary data formats [4, 10]. The lack of uniform data formats has historically hindered data sharing and reuse, as each system generates its own file structure and metadata. Second, *variability in annotation*: the way events (stimuli, responses) and channels are annotated often differ between labs. Inconsistent event labels or missing metadata hinder others attempting to reproduce an analysis, as crucial context may be lost or misinterpreted. Third, *variability in recording conditions*: EEG is collected from diverse populations (developmental, clinical, and healthy adults) and in various tasks. Differences in signal characteristics across populations or tasks may necessitate bespoke preprocessing, rendering a one-size-fits-all pipeline impractical.

## 2.2 Heterogeneity of Analysis Methods and Tools

*Variability in analysis method*: analysts make numerous choices (e.g., filter characteristics, artifact-rejection thresholds, and re-referencing strategies), which can significantly affect outcomes. Because EEG signals are low-amplitude and prone to noise, researchers often differ in how they clean the data. Such flexibility means that even when testing the *same hypothesis*, different analysis pipelines can yield different results [10, 11]. Indeed, a recent large-scale multi-analyst project ("EEGManyPipelines") involved 168 teams analyzing the same EEG dataset and found wide variability in the processing approaches used, which, in turn, led to variability in the reported results [11]. This underscores how degrees of freedom in EEG analysis can threaten robustness.

*Variability in tools*: there are multiple open-source software packages, such as Brainstorm [12, 13], EEGLAB [1, 14], FieldTrip [15], MNE-Python [16], and SPM [17, 18], that provide the same (but often non-identical) implementations of common EEG processing steps. This diversity of tools allows innovation and flexibility. However, it also introduces tool-specific conventions, defaults, and assumptions that can influence analytical outcomes [11].

# 3 Towards Generalizable EEG Analysis

Improving the generalizability of EEG research requires approaches that explicitly acknowledge and address heterogeneity across both EEG datasets and analysis.

## 3.1 Addressing Heterogeneity in EEG Data

To mitigate this issue, the research community has adopted the FAIR principles, which require that data be Findable, Accessible, Interoperable, and Reusable [4, 19, 20]. In EEG, *findability* and *accessibility* refer to datasets that are well documented and hosted in publicly accessible repositories, enabling other researchers to easily discover and use them. *Interoperability* and *reusability* are achieved by using standardized formats and rich metadata, enabling data to be understood and processed by various tools [4, 19].

### The Brain Imaging Data Structure (BIDS)

A landmark development in FAIR principles is the Brain Imaging Data Structure (BIDS), a community-driven standard for organizing neuroimaging data. Initially developed for MRI, BIDS was extended to EEG in 2019 [4]. BIDS-EEG prescribes a consistent folder and filename hierarchy for EEG studies, along with companion metadata files (JSON and TSV sidecars) that record details such as electrode locations, EEG hardware, recording parameters, experimental conditions, and event annotations.

By enforcing a standard file structure, BIDS makes EEG datasets *self-documenting* and tool-agnostic, enabling seamless integration with diverse tools. Critically, BIDS-compliant data contain all the necessary information for analysis in a machine-readable format, enabling fully automated pipelines to run on the data without manual reconfiguration. Moreover, BIDS aligns with FAIR principles: rich metadata and standard nomenclature enhance findability and reusability, and the requirement of open formats (such as EDF and BrainVision) promotes interoperability. Accessibility is facilitated by repositories built on BIDS conventions, such as OpenNeuro [21], which provide cloud storage for shared datasets. The net effect is that an EEG study shared in BIDS format is much easier for others to find, understand, and reuse than one organized ad hoc. This standardization is a foundational step toward more replicable EEG research.

### Hierarchical Event Descriptors (HED)

Accurate interpretation and reuse of EEG data depend critically on clear and standardized descriptions of experimental events. Traditional event annotations based on numeric trigger codes are often study-specific and lack explicit semantic meaning. Hierarchical Event Descriptors (HED) were introduced to address this limitation by providing a structured, machine-readable vocabulary for annotating events in EEG and MEG experiments [22]. HED uses a hierarchical tagging system to describe events by their sensory modality, task context, participant actions, and experimental conditions. These tags can be combined to richly describe an event while remaining interpretable at different levels of abstraction. This approach enables automated querying, consistent event categorization across studies, and improved reproducibility.

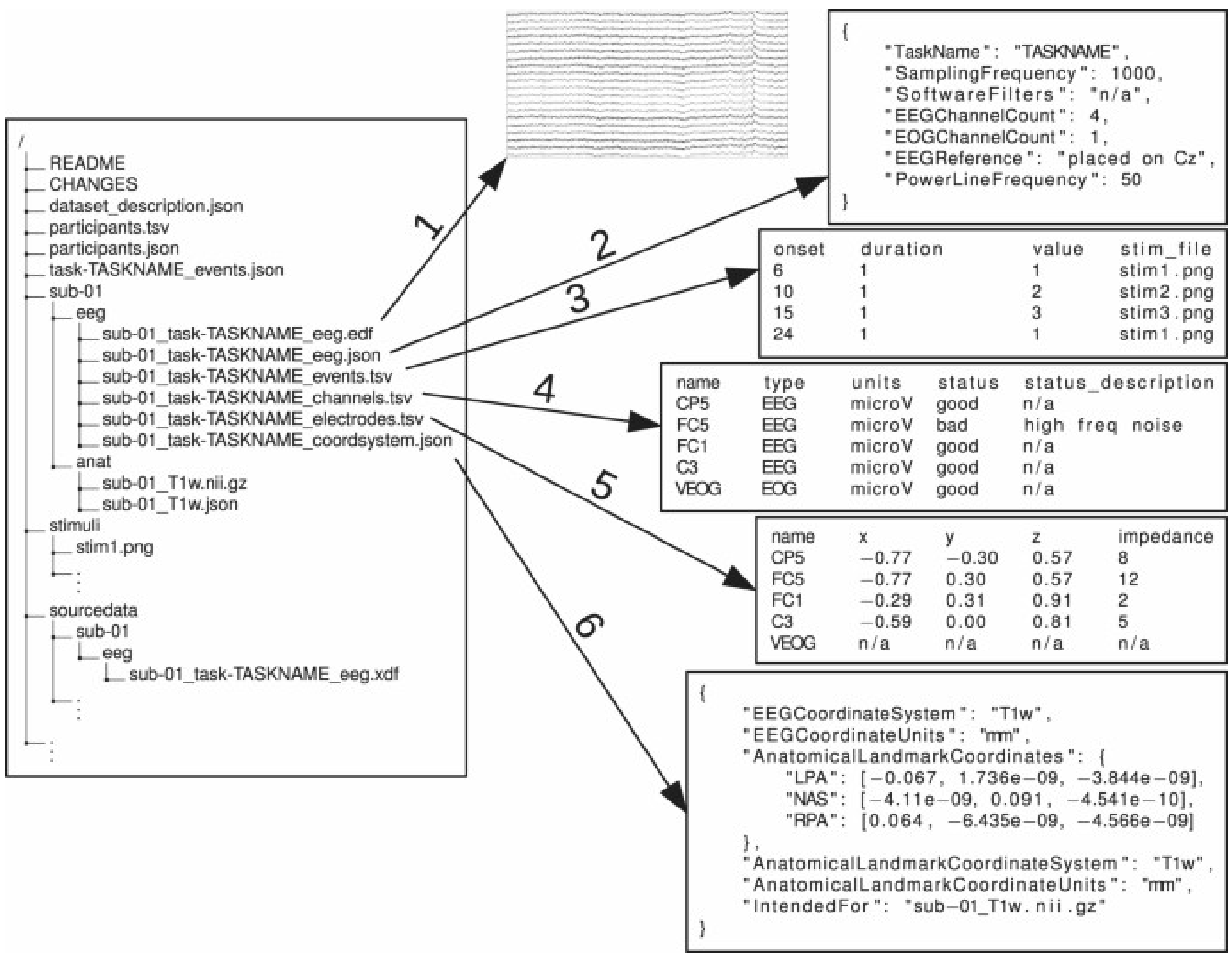

Figure 1: Overview of the BIDS-EEG data organization and associated metadata files. Schematic illustration of a Brain Imaging Data Structure (BIDS) compliant EEG dataset, showing the directory hierarchy and the linkage between raw EEG recordings and their accompanying metadata. Adapted from Pernet et al. (2019), *Scientific Data*, under the Creative Commons Attribution 4.0 International License.

HED is fully integrated within the BIDS format, where tags are stored alongside event timing information in sidecar files. This integration allows datasets to preserve detailed semantic information about experimental events without modifying raw EEG data, facilitating large-scale data sharing, automated pipelines, and cross-study analyses.

### 3.2 Addressing Heterogeneity in Methods and Tools

Variability in EEG analysis methods and software implementations represents a major barrier to generalizable outcomes. Differences in algorithmic choices, parameterization, software defaults, and computational environments can lead to substantively different results, even when analyses are conceptually aligned. In this sense, open-source software is crucial for transparent EEG analysis [23, 24]. If the code for processing data is freely available, others can review and run it, thereby increasing trust and enabling exact reproduction of the analysis. Today, the major software for EEG analysis is open-source and encourages community contributions. This open-source approach not only includes analysis software (discussed in the next section) but also extends to data-sharing platforms (such as OpenNeuro [21]) and standards organizations (such as the International Neuroinformatics Coordinating Facility for BIDS [25]). Community guidelines explicitly recommend sharing both data and analysis pipelines to improve reproducibility. For example, the COBIDAS MEEG committee of the Organization for Human Brain Mapping (OHBM) published best-practice recommendations urging researchers to share detailed processing protocols and code scripts [8]. The culture is shifting towards what might be called "open methods," utilizing collaborative tools (GitHub for code, forums for support), adopting version control for analyses, and prioritizing transparency at every step.

## 4 Major Open-source Software for EEG Analysis

The landscape of EEG analysis software is rich, with each software offering unique features and philosophies. Here, we compare the major open-source software, focusing on how they support reproducible and rigorous EEG research. We consider both MATLAB- and Python-based toolkits, as well as specialized tools, highlighting their strengths, limitations, and best-use contexts.

### 4.1 Brainstorm

**Environment & Interface:** Brainstorm (Tadel et al. 2011) is a free, open-source platform [12, 13] that emphasizes an intuitive graphical user interface (GUI) and powerful visualization capabilities. It runs in MATLAB (with some Java) or as a standalone compiled application (no MATLAB license needed). Data is managed in a *database* organized by subject and condition. Each subject's folder contains raw recordings and any processed results, along with that subject's anatomy (MRI/surface) if applicable. This *protocol-based* structure ensures consistency by automatically tracking which data and results belong together.

**Reproducibility:** Nearly every GUI action in Brainstorm can be converted to a MATLAB script automatically. Users can prototype an analysis interactively, then save and edit the equivalent script to run the workflow on new subjects or replicate it elsewhere. Brainstorm also offers a visual *Pipeline Editor*, where processing steps are graphically linked and executed sequentially. These pipelines can be saved as "process files" and applied to other datasets. In addition, Brainstorm can create a "snapshot" of the current state of the source code and installed plugins, which can be archived and distributed alongside the processing scripts.

**Capabilities:** Brainstorm is particularly strong in source modeling and multimodal integration. It includes advanced head modeling tools, such as the boundary-element and finite-element methods, via OpenMEEG [26], DUNEuro [27], and iso2mesh [28] libraries. EEG and MEG data can be co-registered to individual MRIs, and results can be visualized on 3D cortical surfaces. Brainstorm supports a broad range of analyses, from basic sensor-space ERPs and spectra to source localization, connectivity, and group-level statistics [13, 29].

**BIDS/HED Support:** Brainstorm has built-in support for EEG-BIDS. A one-click importer reads a BIDS dataset and populates the Brainstorm database with subjects and recordings. Likewise, Brainstorm can export cleaned data or source maps along with metadata, facilitating sharing.

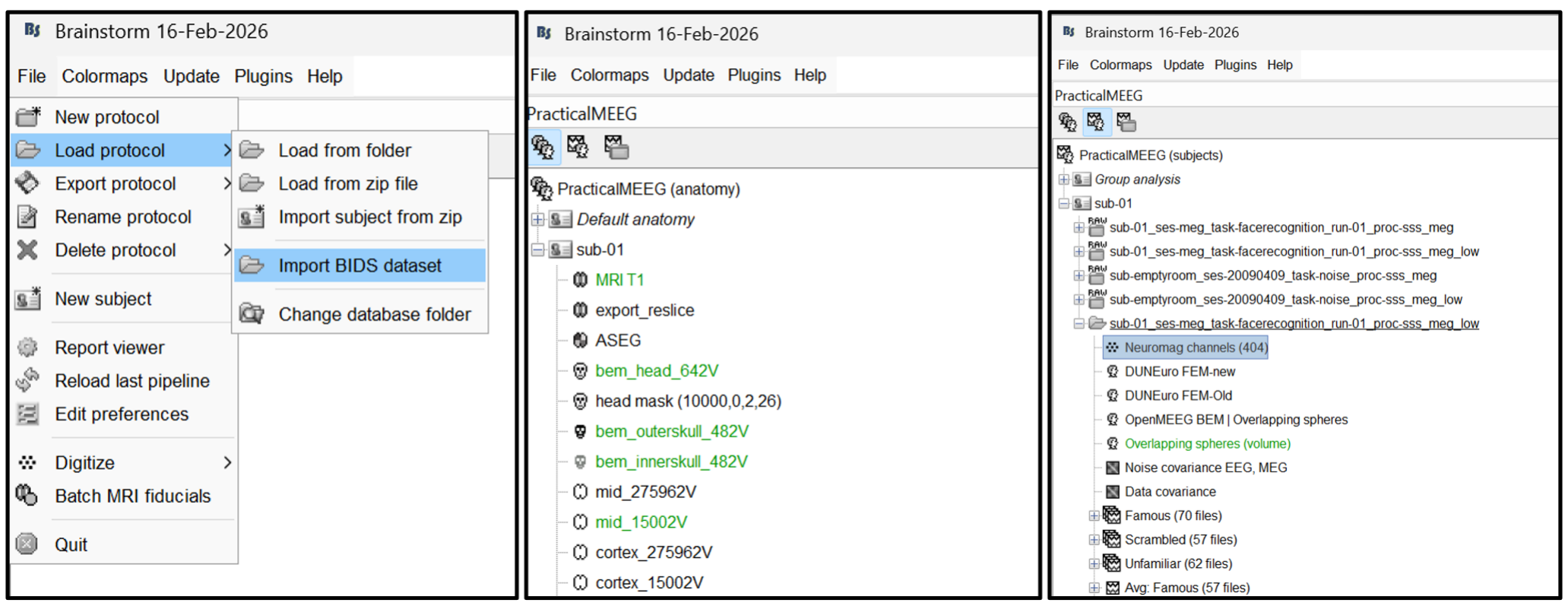

Figure 2: Brainstorm database organization illustrated using the PracticalMEEG dataset. Left: One-step automatic import interface for MEG-BIDS datasets in Brainstorm. Center: Subject-specific anatomical files, including MRI volumes and reconstructed cortical surfaces. Right: Organized MEG/EEG recordings and associated preprocessing and analysis results within the Brainstorm database structure.

**Plugins, Interoperability, and Ecosystem Integration:** Brainstorm is designed as a central hub rather than a closed analysis environment. Its plugin system enables seamless interoperability with a wide range of external neuroimaging tools and libraries. Through this architecture, Brainstorm interfaces with multiple established packages, including OpenMEEG and DUNEuro for forward modeling, FieldTrip [15] for advanced signal processing and connectivity analysis, SPM12 [18] and CAT12 [30] for anatomical processing, BrainSuite [31] for cortical reconstruction, Iso2Mesh [28] and Zeffiro [32] for meshing, and NIRSTORM [33] for near-infrared spectroscopy analysis, among others.

**Strengths & Limitations:** Brainstorm's GUI approach makes it accessible to novices and great for teaching. Its rich sensor- and source-level tools have made it popular among MEG researchers, but EEG analysts also benefit from its ease of use and visualization. The GUI scripting bridge appeals to those concerned with reproducibility. A limitation is that GUI-based flows can be less flexible than coding, and highly customized methods may be harder to implement. However, the active development community and continuous updates keep Brainstorm at the cutting edge.

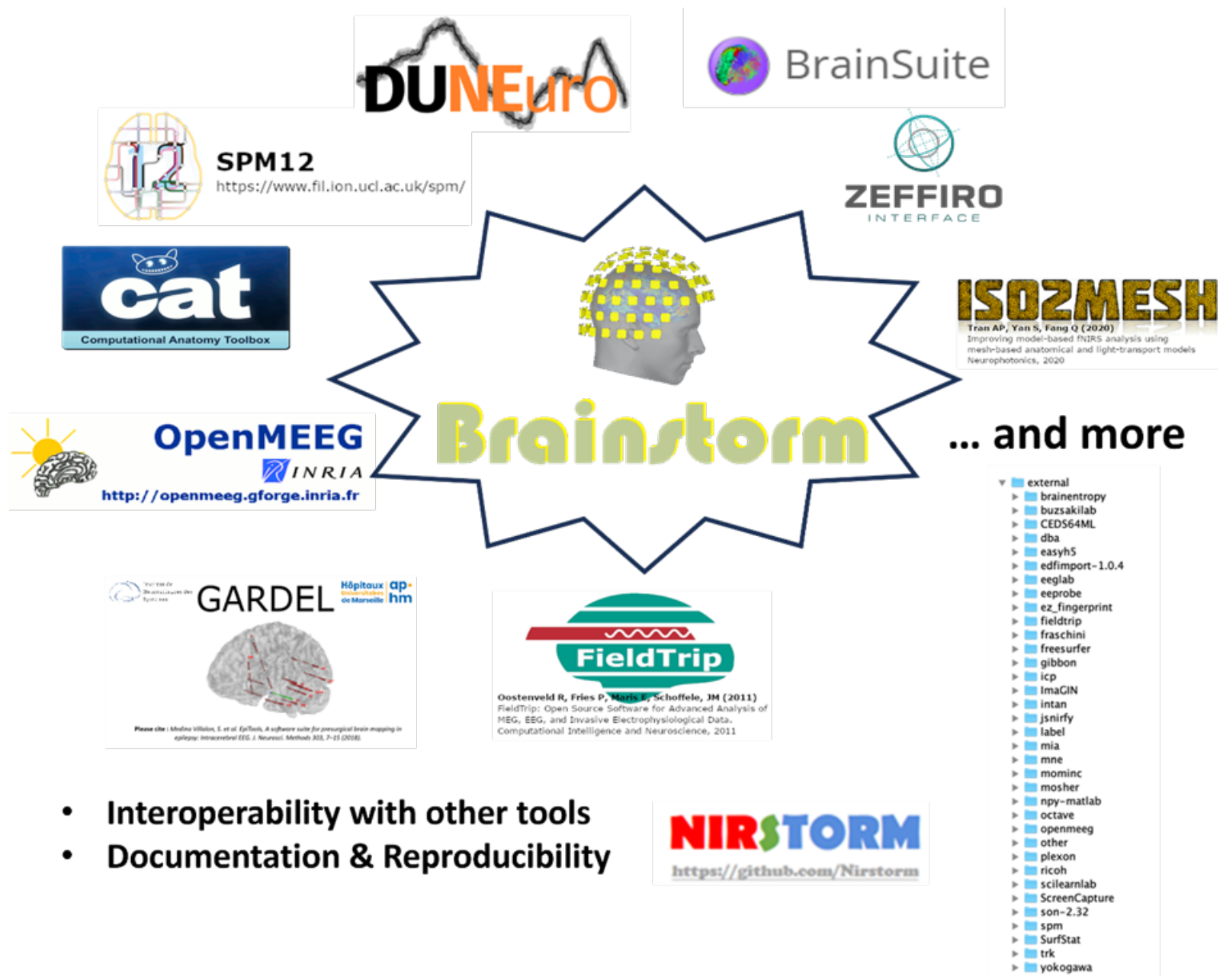

Figure 3: Brainstorm Plugin Manager as an Interoperability Hub. Brainstorm integrates with a broad ecosystem of neuroimaging tools (e.g., SPM, FieldTrip, OpenMEEG, BrainSuite, CAT, and others) through its plugin manager, enabling interoperable workflows, standardized data exchange, and reproducible analysis pipelines.

## 4.2 EEGLAB

**Environment & Interface:** MATLAB-based toolbox with a basic GUI and scripting toolbox (Delorme et al. 2004) [14]. Users can run analyses interactively or via scripts. EEGLAB's design is modular, with hundreds of community plugins extending its capabilities.

**Data Handling:** Supports multiple EEG file formats (e.g., BrainVision, EDF) and provides the STUDY structure to organize data from multiple subjects for group analysis [1]. Crucially, EEGLAB logs GUI actions: every menu-driven operation is recorded as a MATLAB command in the EEG dataset's history field. This logging enables users to reproduce a GUI analysis later or export the processing as script code.

**Standards Support:** EEGLAB offers one-click import of BIDS/HED datasets via community plugins. Conversely, it can export data/results to standard formats. Thus, EEGLAB can be integrated into BIDS-based pipelines.

**Reproducibility:** The STUDY structure enables users to apply the same preprocessing and statistical workflow across subjects [1]. Several EEGLAB plugins automate previously subjective steps, such as the PREP pipeline (line-noise removal and robust referencing) and ICLabel (ML-based ICA component classification) [34], thereby standardizing artifact correction. One study demonstrated a fully reproducible EEGLAB workflow, starting from a publicly available dataset [1, 35].

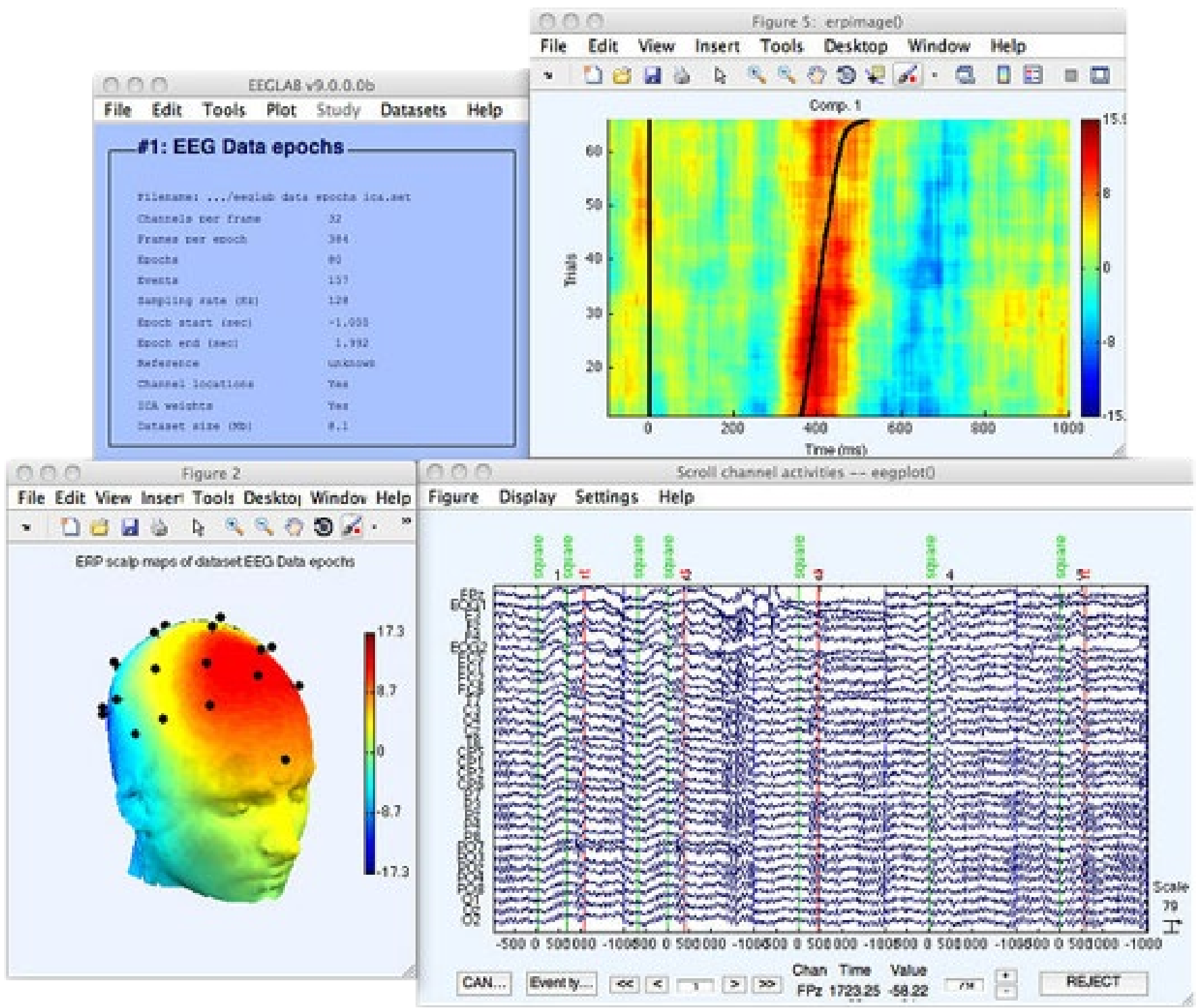

Figure 4: Example EEGLAB graphical user interface and visualization outputs. Screenshots show dataset information, an ERP image plot, scalp topography, and scrolling channel data, illustrating EEG visualization and preprocessing in EEGLAB.

**Strengths & Limitations:** EEGLAB's primary strength lies in its large, active community and highly extensible plugin ecosystem. A wide range of validated plugins supports specialized analyses, including artifact subspace reconstruction (ASR) for automated artifact removal, advanced ICA workflows, time–frequency decomposition, connectivity measures, and machine-learning extensions such as BCILAB [36]. EEGLAB is primarily designed for EEG, with limited native support for MEG, source modeling, or multimodal forward solutions. Notably, the Open EEGLAB Portal [37] allows users to run EEGLAB pipelines on HPC infrastructures via a web interface.

## 4.3 FieldTrip

**Environment & Interface:** FieldTrip is a MATLAB-based toolbox (Oostenveld et al. 2011) [15] that is script-driven. Users write MATLAB scripts calling the ft_ functions for each processing step. FieldTrip emphasizes a transparent pipeline: every function takes a config struct, so all parameters are explicitly set (or left at documented defaults).

```
% start EEGLAB
clear; [ALLEEG, EEG, CURRENTSET, ALLCOM] = eeglab;

% call BIDS tool BIDS
studypath      ='XXX\ds002718';
[STUDY, ALLEEG] = pop_importbids(filepath,'bidsevent','on','bidschanloc','on',...
            'studyName','Face_detection','outputdir', fullfile(filepath, 'derivatives'), ...
            'eventtype', 'trial_type');
ALLEEG        = pop_select( ALLEEG, 'nochannel',{'EEG061','EEG062','EEG063','EEG064'});
CURRENTSTUDY    = 1;
EEG           = ALLEEG;
CURRENTSET      = 1:length(EEG);

% Remove bad channels
EEG = pop_clean_rawdata( EEG,'FlatlineCriterion',5,'ChannelCriterion',0.8,...
   'LineNoiseCriterion',4,'Highpass',[0.25 0.75] ,...
   'BurstCriterion','off','WindowCriterion','off','BurstRejection','off',...
   'Distance','Euclidian','WindowCriterionTolerances','off' );

% Rereference using average reference
EEG = pop_reref( EEG,[],'interpchan',[]);

% Run ICA and flag artifactual components using IClabel
for s=1:size(EEG,2)
   EEG(s) = pop_runica(EEG(s), 'icatype','runica','concatcond','on','options',{'pca',EEG(s).nbchan-1});
   EEG(s) = pop_iclabel(EEG(s),'default');
   EEG(s) = pop_icflag(EEG(s),[NaN NaN;0.8 1;0.8 1;NaN NaN;NaN NaN;NaN NaN;NaN NaN]);
   EEG(s) = pop_subcomp(EEG(s), find(EEG(s).reject.gcompreject), 0);
end

% clear data using ASR - just the bad epochs
EEG = pop_clean_rawdata( EEG,'FlatlineCriterion','off','ChannelCriterion','off',...
   'LineNoiseCriterion','off','Highpass','off','BurstCriterion',20,...
   'WindowCriterion',0.25,'BurstRejection','on','Distance','Euclidian',...
   'WindowCriterionTolerances',[-Inf 7] );

% Extract data epochs (no baseline removed)
EEG    = pop_epoch( EEG,{'famous_new','famous_second_early','famous_second_late', ...
      'scrambled_new','scrambled_second_early','scrambled_second_late','unfamiliar_new', ...
      'unfamiliar_second_early','unfamiliar_second_late'},[-0.5 1] ,'epochinfo','yes');
EEG    = eeg_checkset(EEG);
EEG    = pop_saveset(EEG, 'savemode', 'resave');
ALLEEG = EEG;

% update study & compute single trials
STUDY       = std_checkset(STUDY, ALLEEG);
[STUDY, EEG] = std_precomp(STUDY, EEG, {}, 'savetrials','on','interp','on','recompute','on',...
   'erp','on','erpparams', {'rmbase' [-200 0]}, 'spec','off', 'ersp','off','itc','off');
eeglab redraw
```

Import EEG-BIDS data
remove unwanted channels
create a STUDY

Remove bad channels
Filtering
Re-referencing

ICA decomposition
Component labelling
Artefact rejection

Extract epochs
Computes single trials

Figure 5: Example EEGLAB script for importing and preprocessing an EEG-BIDS dataset. The workflow includes BIDS import, channel cleaning, filtering, re-referencing, independent component analysis (ICA), automated component labeling (ICLabel), artifact rejection, epoch extraction, and computation of single-trial measures within an EEGLAB STUDY structure. The full version of this script is available in the LIMO-EEG repository.

**Capabilities:** FieldTrip is renowned for its flexibility and depth. It supports standard EEG/MEG processing (preprocessing, time–frequency transforms, event-related potentials, connectivity, and statistics), as well as advanced methods (source reconstruction using various inverse techniques, brain–computer interfaces, and real-time processing). It has extensive statistical tools, such as the cluster permutations [38]. For example, Andersen (2018) [39] demonstrated a fully reproducible group time–frequency analysis entirely in FieldTrip.

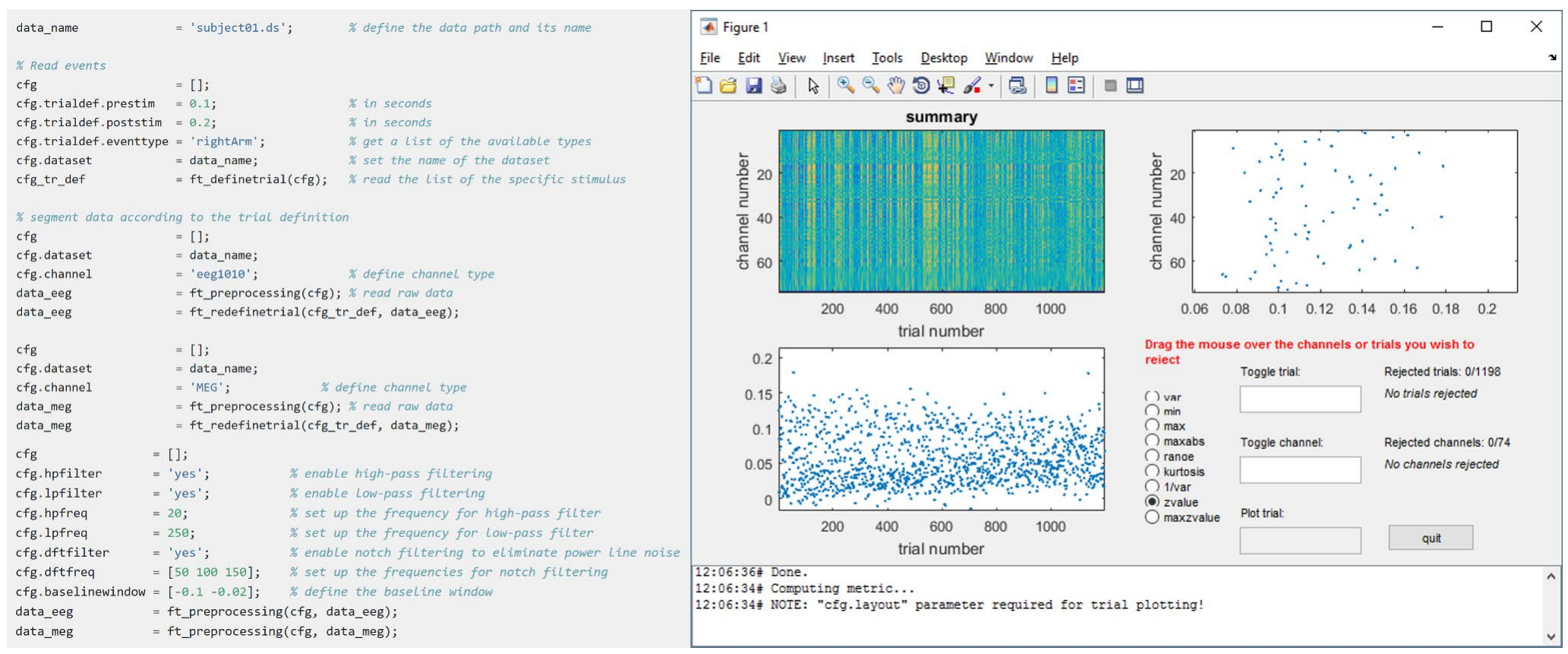

Figure 6: FieldTrip scripting workflow and data inspection interface. Left: An example of a MATLAB script implementing FieldTrip preprocessing steps (trial definition, filtering, and baseline correction). Right: The interactive data browser for visualizing trials and identifying artifacts.

**Standards Support:** FieldTrip does not have a one-click BIDS importer, but it is largely "BIDS-ready". It can read raw EEG data in common formats using ft_read_raw and related functions. Community examples show how to loop over a BIDS directory structure and call FieldTrip readers for each run.

**Reproducibility:** Because FieldTrip analysis is entirely scripted, reproducibility is guaranteed if the script and FieldTrip version are shared. FieldTrip's integration with other toolboxes (EEGLAB, Brainstorm) allows cross-validation.

**Strengths & Limitations:** FieldTrip's principal strength lies in the level of analytical control it offers and the breadth of methods it provides. It is particularly well-suited for researchers who require access to advanced and cutting-edge techniques, especially in source analysis and functional connectivity. It includes extensive support for source localization approaches, such as beamforming and distributed minimum-norm methods [40]. This flexibility comes with a steeper learning curve. FieldTrip does not provide a graphical user interface, and analyses are performed entirely through MATLAB scripting. At the same time, FieldTrip's transparency strongly supports reproducibility when good practices are followed [41]. FieldTrip also integrates well with other MATLAB-based toolboxes through dedicated data-conversion utilities, enabling cross-tool validation [42].

## 4.4 MNE-Python

**Environment & Interface:** A comprehensive Python library for EEG/MEG (Gramfort et al.) [16] built on NumPy/SciPy. MNE uses a scripted approach (Python scripts or Jupyter notebooks) for all analyses. The library includes a *Report* class that automatically collects figures and generates an HTML report summarizing each analysis.

**Capabilities:** MNE-Python supports all major EEG/MEG steps, including raw data I/O and preprocessing, filtering, ICA artifact removal, time–frequency analysis, source localization, connectivity, and statistics. It also integrates machine learning, enabling extraction of features for use with scikit-learn [43] or deep learning libraries.

**Standards Support:** MNE was a pioneer in EEG-BIDS; the mne-bids package enables users to read BIDS-formatted EEG/MEG datasets with just a few lines of code. There is also a standardized MNE-BIDS pipeline distributed as a BIDS App [44].

**Reproducibility:** Because MNE-Python analyses are code-driven, reproducibility is high when scripts and environment are shared. The Python ecosystem (Conda, pip) makes it easy to capture exact software versions; one can use Conda or Docker so that everyone uses the same versions of MNE, NumPy, and other libraries.

**Strengths & Limitations:** MNE-Python's advantages include its free Python base (no MATLAB required), integration with modern data science tools, and a strong community. It is ideal for users who want to integrate EEG analysis with other data (e.g., machine learning and deep learning). MNE-Python is widely used in both academia and industry because Python's ecosystem makes it easy to connect EEG analysis with web services, databases, and machine learning. The transition to cloud and HPC is also well supported.

## 4.5 SPM (Statistical Parametric Mapping)

**Environment and Interface:** SPM (Litvak et al. 2011) is a MATLAB-based software package originally developed for the statistical analysis of neuroimaging data [17], with long-standing support for EEG and MEG. It provides both a graphical user interface and a scripting API. EEG/MEG data are organized within a structured directory hierarchy, and analyses are configured through MATLAB "batch" objects that explicitly define each processing step.

**Capabilities:** SPM offers a comprehensive suite of tools for EEG and MEG analysis, with particular strength in source reconstruction and statistically rigorous inference. It is especially well known for its Bayesian generative modeling framework, including dynamic causal modeling (DCM) for investigating effective connectivity [45]. These features make SPM particularly attractive for hypothesis-driven modeling and group-level statistical inference [46].

**Standards Support:** SPM supports reading common EEG and MEG data formats. While SPM does not provide a one-click EEG-BIDS importer, it can be integrated into BIDS-based workflows through conversion scripts and external tools.

**Reproducibility:** SPM's batch system strongly supports reproducibility. Each GUI-based analysis corresponds to a MATLAB batch structure that can be saved, version-controlled, and re-executed on new datasets. SPM maintains detailed provenance information within its output files.

**Strengths and Limitations:** SPM's principal strength lies in its statistically principled approach, particularly at the source and group levels. Its Bayesian modeling framework and tools [47], such as DCM, are widely used to test mechanistic hypotheses about brain function. Limitations include a steeper conceptual learning curve and less extensive sensor-level preprocessing compared to EEGLAB or Brainstorm. In practice, SPM is frequently combined with other ecosystems, forming part of a broader interoperable workflow.

# 5 Established and Automated Preprocessing Pipelines

Many automated preprocessing pipelines have been developed. Here we highlight a few of them, as the list is still growing and improving every day.

## 5.1 General Pipelines

**PREP:** An early standardized pipeline [48] that focuses on identifying bad channels and computing a robust average reference. PREP removes line noise adaptively and generates a comprehensive report, ensuring uniform preprocessing across datasets.

**ADJUST:** An automated ICA-based artifact detector [49] that identifies stereotypical artifacts (eye blinks, eye movements, discontinuities) by spatial and temporal features. ADJUST classifies ICA components without manual marking.

**MARA:** The Multiple Artifact Rejection Algorithm [50, 51] is a machine learning classifier that assigns an artifact probability to each ICA component. Components with a high probability are automatically removed.

**FASTER:** A fully automated pipeline [52] that applies statistical thresholds to the EEG. It computes metrics for each channel or epoch (e.g., amplitude, variance, and correlation) and automatically interpolates or removes values that exceed normalized thresholds.

**Automagic:** A MATLAB toolbox [53] that integrates PREP and several other methods into one turnkey pipeline. It performs robust referencing, ICA, and artifact subspace reconstruction via a graphical interface.

**RELAX:** A recent fully-automated EEGLAB plugin for EEG cleaning [54]. It applies multiple algorithms, including ICA and artifact subspace analysis.

**BrainFusion:** A recent low-code framework for multimodal neurophysiological analysis [55], providing standardized and reproducible preprocessing pipelines across various signals, including EEG, fNIRS, EMG, and ECG.

These tools shift artifact correction from manual, researcher-driven decisions to standardized procedures, reducing "researcher degrees of freedom." Each records its actions in log files or reports, making it clear what was done to the data.

## 5.2 Developmental Pipelines

EEG from infants and children poses extra challenges (motion artifacts, evolving head size, and EEG properties). Specialized pipelines automate preprocessing for developmental data:

**HAPPE:** The Harvard Automated Processing Pipeline for EEG [56] is designed for high-artifact developmental EEG. HAPPE first applies wavelet-enhanced ICA to remove severe artifacts, then uses MARA to reject ICA components.

**MADE:** Multivariate Artifact Detection for EEG [57], a pipeline developed for infant EEG emphasizing minimal trial rejection and preservation of signal.

**APICE:** Automated Pipeline for Infants Continuous EEG [58], a flexible EEGLAB-based pipeline. APICE is fully automated and modular, applying multiple adaptive algorithms on continuous data.

Additional pipelines such as PREP+ER, BEAPP, and NEAR have been proposed for developmental EEG.

## 5.3 Source Imaging Pipelines

For studies involving source reconstruction and connectivity, standardized pipelines such as CiftiStorm have been introduced [59]. CiftiStorm is an end-to-end electrophysiological source-imaging pipeline that incorporates cutting-edge forward/inverse techniques. It produces outputs in CIFTI format (the standard format of the Human Connectome Project) from input datasets of varying sensor densities. Its outputs are HCP-compliant, facilitating multimodal connectomics.

Across all these pipelines, the trend is toward automation, as reducing manual intervention in preprocessing improves transparency and enables easy rerunning or comparison of analyses.

Table 2: Overview of EEG analysis, acquisition, and processing software ecosystems. Representative tools are organized by functional domain, highlighting the diversity of tools supporting reproducible EEG research workflows.

| Category | Tool | Key Notes / Use-case |
|---|---|---|
| **Commercial Software** | BESA | Full-featured, GUI-driven; source localization |
| | LabChart Lightning | Good for EEG/EMG/ECG acquisition |
| | encevis | Medically certified; clinical diagnostics |
| | Persyst | Clinical EEG; automated spike detection and AI-assisted diagnostics |
| | BrainVision Analyzer | MATLAB-based, offline chain with rich feature set |
| | BrainVision RecView | Real-time analysis (e.g., for BCI) |
| | EGI NetStation | Integrated acquisition and analysis; high-density and developmental EEG |
| | AcqKnowledge | Tied to BIOPAC hardware |
| | CURRY (Compumedics Neuroscan) | EEG/MEG source localization, forward/inverse modeling |
| **LSL Integrations / Acquisition Frameworks** | LabRecorder | Recording synchronized data streams via LSL |
| | NeuroPype | Real-time BCI and signal processing pipelines |
| | OpenViBE | Open-source platform for BCI design and testing |
| | BrainFlow | Device-agnostic API; hardware abstraction layer |
| **MATLAB-based Tools** | EEGLAB | Popular toolbox; many plugins; strong for EEG/MEG workflows |

*continued on next page*

Table 2 – *continued from previous page*

| Category | Tool | Key Notes / Use-case |
|---|---|---|
| | PREP | Standardized preprocessing; robust referencing and line-noise removal |
| | RELAX | Automated EEG cleaning; ICA-based artifact rejection |
| | ERPLab | Focused on ERP analyses; works with EEGLAB |
| | Mass Univariate ERP Toolbox | Mass-univariate analyses of ERPs |
| | FieldTrip | Broad EEG/MEG/iEEG functionality; source reconstruction |
| | BCILAB | For BCI research, feature extraction, and machine learning |
| | BioSig | Library for many biosignals |
| | SPM | Neuroimaging suite; Bayesian stats and group-level inference |
| | NutMEG | MEG/EEG source localization |
| | Chronux | Time-series analysis, spectral estimation |
| | WU-NEAT | Validated neonatal EEG toolbox |
| **Python-based Tools** | MNE-Python | Full suite for EEG/MEG/iEEG/fNIRS |
| | PsychoPy | Experiments and behavioral paradigms |
| | PyEEG | Time-series feature extraction |
| | pyEDFlib | Reading/writing EDF/BDF files |
| | NeuroKit2 | Processing neurophysiological signals (EEG, ECG, EDA, etc.) |
| | FOOOF | Spectral parameterization (periodic vs. aperiodic) |
| | osl-ephys | Extensions to MNE-Python for large-scale M/EEG |
| | Autoreject | Automated bad-trial rejection/repair |
| | HyPyP | Python hyperscanning pipeline |
| **Standalone / Multimodal Platforms** | Brainstorm | Open-source; standalone UI; rich multimodal analysis |
| | Cartool | C++ based; visualization, inverse solutions |
| **Application-specific Pipelines** | HAPPE | Developmental/pediatric EEG |
| | BeMoBIL | Mobile EEG + body imaging |
| | APICE | Automated pipeline for infant EEG |

# 6 Community Infrastructure and Software Support for BIDS-Compliant EEG Research

The adoption of EEG-BIDS has been driven not only by the specification itself, but also by a rich ecosystem of community-developed tools that support validation, curation, analysis, and annotation. Together, these resources help transform BIDS from a data-organization standard into a practical foundation for reproducible and scalable EEG research.

## 6.1 Validation, Onboarding, and Dataset Curation

At the most basic level, researchers can verify the structural and metadata integrity of EEG-BIDS datasets using the BIDS Validator (https://bids-standard.github.io/bids-validator), which identifies missing files, underspecified metadata, and violations of the standard before sharing or analyzing the data. Beyond validation, dataset creation and onboarding represent a key practical step. Tools such as BIDScoin [60] provide a user-friendly workflow that facilitates mapping raw source data to the BIDS specification. For users new to BIDS, the BIDS Starter Kit (https://github.com/bids-standard/bids-starter-kit) further complements this process with practical tutorials, example workflows, and helper scripts. As datasets grow in size and complexity, tools such as CuBIDS [61] address the gap by supporting systematic dataset curation, enabling researchers to detect hidden variability and harmonize metadata before running automated pipelines.

## 6.2 Scalable and Automated Analysis Pipelines

Recent efforts increasingly treat BIDS as an enabling layer for automated, cohort-level analysis. Frameworks such as BIDS Manager-Pipeline [62] build on BIDS datasets to coordinate multi-subject processing across heterogeneous software tools, while enforcing consistent parameterization and storing outputs using the BIDS derivatives specification. When combined with cloud or high-performance computing resources, these approaches support scalable analyses in which reproducibility is embedded directly into the workflow.

## 6.3 Standardized Event Annotation and Metadata

Standardized event annotation represents a complementary pillar of reproducible EEG research. The HED framework provides a structured, machine-actionable vocabulary for describing experimental events, enabling consistent interpretation of event information across studies and software platforms. To facilitate this process, CTAGGER [63, 64] is proposed as an open-source HED-based annotation tool that can be used either as a standalone application or integrated directly into Brainstorm and EEGLAB workflows. By supporting the creation, inspection, and verification of event annotations, CTAGGER helps ensure that event metadata remains consistent, reusable, and aligned with the broader BIDS ecosystem.

# 7 Recommendations for Good Scientific Practice and Data Transparency

**Adopt Data Standards (BIDS):** Use the Brain Imaging Data Structure (BIDS) to organize EEG data. BIDS prescribes a folder structure and metadata convention that makes datasets self-describing and machine-readable. EEG-BIDS [1] extends this to EEG, promoting FAIR principles. Standards bodies (OHBM, COBIDAS, MEEG [8]) explicitly encourage the use of BIDS. Most toolkits now support EEG-BIDS. Using BIDS, data can be shared in repositories (e.g., OpenNeuro) and reused via automated pipelines [7].

**Share Code and Environment:** Always report and share all analysis code (scripts or notebooks) used. Specify software versions explicitly. Python users can use Conda or Container Tools such as Docker to ensure that specific library versions are captured. Lab notebooks or READMEs should list MATLAB/toolbox versions, as well as plugin versions. In the case of Brainstorm, it provides the tools to archive the current Brainstorm environment and its plugins, so this can be shared alongside the analysis code.

**Pre-registration and Transparency:** When conducting hypothesis-driven EEG studies, pre-register the analysis plan to ensure transparency and reproducibility. Pre-registration (e.g., via the Open Science Framework or as *Registered Reports*) fixes the hypotheses, planned preprocessing, and statistical tests *before* seeing the data. This combats "analytic flexibility." The EEGManyLabs [10] and EEGManyPipelines [11] consortia submit Registered Reports of their replication designs.

**Automated Quality Control:** Implement objective quality checks using automated tools. For example, the *Autoreject* algorithm [65] automatically detects bad epochs or channels and interpolates or rejects them, with minimal human intervention. Automated QC tools produce logs of what was fixed, enhancing transparency. Many BIDS apps and pipelines include QC reports to flag problems.

**Multiverse and Robustness Checks:** Since many valid preprocessing choices exist, it is good practice to test the robustness of results across pipeline variants. One approach is a "multiverse" analysis: run several plausible pipelines (e.g., different filter settings, ICA vs. subspace methods) and examine consistency of outcomes. Comparing outputs across toolkits (e.g., EEGLAB vs. FieldTrip) on the same data can also reveal an undue reliance on software details.

**Community Guidelines (COBIDAS, etc.):** Follow established recommendations. The OHBM COBIDAS MEEG report provides a checklist of items to report (e.g., hardware details, preprocessing steps, software versions). Adhering to these guidelines ensures that others have all the necessary information to reproduce or reanalyze your work.